\documentclass[aps,pra,reprint, amsmath, amssymb,superscriptaddress]{revtex4-1}

\usepackage{bm}
\usepackage[retainorgcmds]{IEEEtrantools}
\usepackage{graphicx}
\usepackage{mathrsfs}
\usepackage{amsmath}
\usepackage{amssymb}
\usepackage{color}
\usepackage{amsfonts}
\usepackage{times,txfonts}
\usepackage{nicefrac}
\usepackage{physics}

\newcommand{\trans}{^{\text{T}}}
\newcommand{\inv}{^{-1}}

\newcommand{\eq}[1]{(\ref{#1})}
\newcommand{\ii}{\mathrm{i}}

\begin{document}

\title{
On the Strong Subadditivity of the R\'enyi entropies for bosonic and fermionic Gaussian states}
\date{\today}
\author{Giancarlo Camilo}
\email{gcamilo@iip.ufrn.br}
\affiliation{International Institute of Physics, Universidade Federal do Rio Grande do Norte, Campus Universit\'ario, Lagoa Nova, Natal-RN 59078-970, Brazil}
\author{Gabriel T. Landi}
\email{gtlandi@if.usp.br}
\affiliation{Instituto de F\'isica da Universidade de S\~ao Paulo,  05314-970 S\~ao Paulo, Brazil}
\author{Sebas Eli\"ens}
\email{seliens@iip.ufrn.br}
\affiliation{International Institute of Physics, Universidade Federal do Rio Grande do Norte, Campus Universit\'ario, Lagoa Nova, Natal-RN 59078-970, Brazil}
\begin{abstract}

Recently, there has been a surge of interest in using R\'enyi entropies as quantifiers of correlations in many-body quantum systems. However, it is well known that in general these entropies do not satisfy the strong subadditivity inequality, which is a central property ensuring the positivity of correlation measures. In fact, in many cases they do not even satisfy the weaker condition of subadditivity.  
In the present paper we shed light on this subject by providing a detailed survey of R\'enyi entropies for bosonic and fermionic Gaussian states. We show that for bosons the R\'enyi entropies always satisfy subadditivity, but not necessarily strong subadditivity. 
Conversely, for fermions both do not hold in general. 
We provide the precise intervals of the R\'enyi index $\alpha$ for which subadditivity and strong subadditivity are valid in each case.

\end{abstract}
\maketitle{}

%
%
\section{\label{sec:int}Introduction}
%
%

There is currently a large effort from the quantum physics community to further our understanding of the typical correlation patterns of many-body quantum states \cite{Huber2010,Bennett2011,Levi2013,Schwaiger2015,Goold2015a,Girolami2017,Calabrese2005}. Examples include the dynamics of quantum quenches \cite{calabrese_evolution_2005,calabrese_entanglement_2007,2011_Santos_PRL_107,nezhadhaghighi_entanglement_2014,alba_quench_2017,alba_entanglement_2018,alba_renyi_2017}, quantum Markov chains \cite{Kato2016,Renner2002,Ibinson2008}, Bell non-locality \cite{Chaves2016}, among others. 
Remarkably, correlation patterns are also starting to become accessible to controlled quantum platforms. 
For instance, the correlations in a 8-site  Bose-Hubbard model have been measured in~\cite{Islam2015}, while in Ref.~\cite{Lanyon2017} the authors have implemented a type of matrix product state tomography for a trapped ion system.
These efforts are helping to shed light on the important question of typicality of many body states \cite{Bianchi2018,Eisert2010a,Alba2017,Calabrese2005,Eisler2007,Zurek2018,Deutsch2018,Srednicki1994,Deutsch1991,Kaufman2016,Popescu2006,White1992,Schollwock2011,DeChiara2006,McCulloch2007}, i.e.,  which are the typical sectors of the Hilbert space that are usually occupied by systems with well defined structures. 

Central to this discussion, therefore, are the tools used to quantify correlations. 
Quite often, these are based on  entropic quantities. 
In the simplest scenario, a system with density matrix $\rho_{AB}$ is divided into a bipartition $AB$. 
The amount of information shared between the two parts is then characterized by the Mutual Information (MI) defined as  \cite{Nielsen}
\begin{equation}\label{MI}
\mathcal{I}(A\!:\!B) = S(\rho_A) + S(\rho_B) - S(\rho_{AB}) \geq 0\,,
\end{equation}
where $S(\rho) = - \tr(\rho \ln \rho)$ is the von Neumann entropy of $\rho$ and $\rho_A = \tr_B \rho_{AB}$ and $\rho_B = \tr_A \rho_{AB}$ are the reduced density matrices of $A$ and $B$.
The MI is always non-negative, a fact known as the \emph{subadditivity} (SA) of the von Neumann entropy. Moreover, it is zero if and only if $A$ and $B$ are in a product state, i.e., $\rho_{AB}=\rho_A\otimes\rho_B$.
This ensures that $\mathcal{I}(A\!:\!B)$ is a genuine measure of correlations. 
When $\rho_{AB}$ is a pure state we get $S(\rho_{AB}) = 0$ and $S(\rho_A) = S(\rho_B)$. In this case the MI becomes twice the entanglement entropy. 
Instead, for mixed states the MI provides a measure of the total amount of correlations (quantum and classical) between $A$ and $B$. 

In addition to subadditivity, the von Neumann entropy also satisfies a more general inequality known as strong subadditivity (SSA) \cite{Nielsen}. 
Namely, given a tripartite state $\rho_{ABC}$, then
\begin{IEEEeqnarray}{rCl}\label{SSA}
S(\rho_{AB}) + S(\rho_{BC}) &\geq&  S(\rho_{ABC}) + S(\rho_B)\,.
\end{IEEEeqnarray}
The proof of this inequality, which is rather easy in the classical setting, turns out to be much more complicated in the quantum scenario \cite{Lieb1973} (although for theories with a holographic dual it is remarkably simple \cite{Headrick:2007km}). 
The SSA is one of the most fundamental results in quantum information theory, being the bedrock behind a large number of important results, including quantification of multipartite correlations. 
Indeed, from it one is  naturally led to define a quantity called the  Conditional Mutual Information (CMI) \cite{Renner2002,Ibinson2008,Kato2016,Fawzi,Berta2018,Wyner1978}
\begin{equation}\label{CMI}
\mathcal{I}(A\!:\!C|B) = S(\rho_{AB}) + S(\rho_{BC}) - S(\rho_{ABC}) - S(\rho_B) \geq 0\,,
\end{equation}
which quantifies the amount of information shared between $A$ and $C$ \emph{given} knowledge of $B$. 

Despite the enormous success of the von Neumann entropy, in recent years there has been a surge of interest in alternative entropic quantifiers, particularly those given by the so-called R\'enyi entropies, defined as 
\begin{equation}\label{renyi}
S_\alpha(\rho) = \frac{1}{1-\alpha} \ln \tr\rho^\alpha\,,
\end{equation}
where $\alpha \in [0,\infty)$ is a continuous parameter. 
The von Neumann entropy is recovered for $\alpha=1$ (understood as the limit $\alpha\to1$) whereas for $\alpha = 2$ one gets the log of the purity of the state, $S_2(\rho) = - \ln \tr\rho^2$. 
The R\'enyi entropy satisfies several properties expected from an entropic quantifier, such as non-negativity (vanishing for pure states and positive otherwise) and additivity under tensor product. 
However, there are other desired properties that they generally violate. The most important is precisely the SSA \eq{SSA}, which is \emph{not} satisfied for any $\alpha\ne1$. In fact, as shown in Ref.~\cite{Matus2007}, the R\'enyi entropies of a fixed order $\alpha$ in general satisfy no linear inequalities whatsoever (although a \lq\lq weak subadditivity\rq\rq~relation mixing $S_\alpha$ and $S_0$ was proved in \cite{2002quant.ph..4093V} for all $\alpha$). 
As a result, if one na\"ively introduces the R\'enyi Mutual Information (RMI) and R\'enyi Conditional Mutual Information (RCMI) as in \eq{MI} and \eq{CMI}, i.e.,
\begin{align}
\mathcal{I}_\alpha(A\!:\!B) &= S_\alpha(\rho_{A}) + S_\alpha(\rho_{B}) - S_\alpha(\rho_{AB}),\label{RMI}\\[0.3cm]
\mathcal{I}_\alpha(A\!:\!C|B) &= S_\alpha(\rho_{AB}) + S_\alpha(\rho_{BC}) - S_\alpha(\rho_{ABC}) - S_\alpha(\rho_B)\,,\label{RCMI}
\end{align}
these quantities are not ensured to be non-negative and therefore become meaningless as correlation measures.
Explicit violations of SSA were illustrated, for instance, in Ref.~\cite{Adesso2003} for two qubits and in Ref.~\cite{Kormos2017} in the quench dynamics of the transverse field Ising model. 
In spite of this, the quantities above were recently measured experimentally in Ref.~\cite{Islam2015}.


However, quite surprisingly, Adesso \emph{et. al.} have recently found a specific situation that offers a curious exception to the problem mentioned above \cite{Adesso2012} (see also \cite{Adesso:2016ldo}). 
Namely, when restricted to Gaussian states of a many-body bosonic system the SSA turns out to be true for $\alpha = 2$. 
This result offered an interesting alternative for quantifying correlations using $\mathcal{I}_2$ in bosonic Gaussian states, which are of central interest, for instance, to quantum optics and continuous variable quantum computation. It has since led to more theoretical developments including results that have no counterpart for the Von Neumann entropy \cite{lami_log-determinant_2017,lami_schur_2016}.

Apart from this success story of the R\'enyi-2 entropy for bosonic Gaussian states not much is known about the R\'enyi entropies for Gaussian states in general. In particular, there is no information theory for fermionic Gaussian states based on R\'enyi entropies with $\alpha \neq 1$. It is worth stressing at this point that Gaussian states, both bosonic and fermionic, are quite prevalent in modern condensed matter physics, appearing in a multitude of paradigmatic models such as the transverse field Ising model \cite{1970_Barouch_PRA_2,2008_Fagotti_PRA_78,2011_Calabrese_PRL_106,2013_Fagott_PRB_87} and models for topological phases \cite{2006_Kitaev_AP_321,2007_Pachos_AP_322,2007_Feng_PRL_98,2009_Kells_PRB_80,2015_Dubail_PRB_92}. For the transverse field Ising and similar models mappable to free fermions, Gaussian states naturally arise as steady states when the system is brought out of equilibrium by a quantum quench, in accordance with the logic of the generalized Gibbs ensemble, even when the initial state is not Gaussian \cite{2008_Cramer_PRL_100,2013_Fagott_PRB_87,2016_Gluza_PRL_117,2018_Murthy_arxiv,2018_Gluza_arxiv} (note the exception of massless relativistic bosons \cite{sotiriadis_memory-preserving_2016}). Moreover, in a recent work, the dynamics of the logarithmic negativity in such a quench scenario has been related to the RMI with $\alpha = 1/2$ \cite{Alba:2018hie}. 
A moments thought reveals that Gaussian states are also at play behind the scenes in certain mean-field approximations such as Hartree-Fock and Bogoliubov theory.
With the ubiquitousness of Gaussian states and the continued progress in uncovering relations between observables and  entropic quantities \cite{2010PhRvL.104o7201H,Cornfeld:2018sac} in mind, we see that furthering our understanding of SA and SSA for R\'enyi entropies of Gaussian states is both a natural and timely question. 
This is precisely the goal of this work. 

In this paper we set out to map the full range of values $\alpha$ for which the SA and SSA conditions are satisfied in the case of bosonic and fermionic Gaussian many-body states. 
We begin by discussing the covariance matrix approach for computing the R\'enyi entropies for Gaussian states of bosons and fermions in Sec. \ref{sec:entropies}. We make an effort to emphasize as much as possible the similarity between both cases. 
From our development we then find the following results: 
for bosonic Gaussian states we prove that SA is satisfied for all $0 \leq \alpha < \infty$, while for fermions this only holds true for the interval $0 \leq \alpha \leq 2$ and we explicitly give a procedure to construct violations for $\alpha >2$ (Sec. \ref{sec:SA}). Sec. \ref{sec:SSA} is devoted to SSA, where we rely on numerical evidence to conjecture that for bosons SSA holds true in the domain $0\le\alpha \leq 2$ and we show explicit violations for $\alpha > 2$. For fermions, on the other hand, we suggest that SSA holds for $0\le \alpha \leq 1$, although explicit violations in this case are only found for $\alpha \gtrsim 1.3$. In Sec. \ref{sec:final} we gather some concluding remarks.


%
%
\section{\label{sec:entropies}R\'enyi entropy for Gaussian states}
%
%

In this section we discuss how to compute R\'enyi entropies for Gaussian states by considering first bosonic and then fermionic states

\subsection{Bosonic systems}

We consider a system of $N$ bosonic modes $a_1, \ldots, a_N$ satisfying $[a_i,a_j^\dagger] = \delta_{ij}$ and $[a_i, a_j] = 0$. 
We then define the quadrature operators 
\begin{equation}\label{quadrature_bosons}
q_i = \frac{1}{\sqrt{2}} (a_i^\dagger + a_i),
\qquad
p_i = \frac{\ii}{\sqrt{2}}(a_i^\dagger - a_i)
\end{equation}
and collect them in the vector $\bm{X} = (q_1, p_1, \ldots, q_N, p_N)$. 
The canonical commutation relations in terms of $\bm{X}$ are then written as 
\begin{equation}\label{algebra_bosons}
[X_I,X_J] = \ii\, \Omega_{IJ}, 
\qquad 
\Omega = \mathbb{I}_N \otimes (\ii\sigma_y),
\end{equation}
where $I,J=1,\ldots,2N$ and $\sigma_y$ is the Pauli matrix.
The antisymmetric matrix $\Omega$ is the symplectic form of the bosonic algebra \cite{Simon1994,Dutta1995}.

Given a state $\rho$, we now define the $2N\times 2N$ Covariance Matrix (CM) associated to the operators $\bm{X}$ as \cite{PirlaRMP,Adesso2014}
\begin{equation}\label{boson_CM}
\Gamma_{IJ} = \langle \{X_I,X_J\} \rangle,
\end{equation}
where $\{, \}$ represents the anti-commutator and, for simplicity, we assume $\langle X_I \rangle = 0$ since local unitary transformations are not expected to affect the entanglement properties of the state. 
The algebra~(\ref{algebra_bosons}) imposes that any physically reasonable CM must satisfy the following \emph{bona fide} condition
\begin{equation}\label{bona_fide_boson}
\Gamma - \ii\, \Omega \geq 0\,,
\end{equation}
which can be viewed as the generalized Schr\"odinger-Robertson uncertainty relation.

It is well-known from Williamson's theorem \cite{Simon1994,Dutta1995} that any CM may be diagonalized by a symplectic transformation $M$ (i.e., $M \Omega M\trans = \Omega$) so that \eq{algebra_bosons} is preserved and 
\begin{equation}\label{CMdiagonal_boson}
M \Gamma M\trans = \text{diag}(\sigma_1, \sigma_1, \ldots, \sigma_L, \sigma_L)\,,
\end{equation}
where the $\sigma_i \geq 1$ are called the symplectic eigenvalues of $\Gamma$ and correspond to the $N$ positive eigenvalues of $\ii\Omega \Gamma$. It is important to stress that they are not the true eigenvalues of the CM (these are not preserved by $M$), even though their product $\prod_{i=1}^N\sigma_i^2=\det(M\Gamma M\trans)=\det(\Gamma)$ happens to be basis-independent. 
So far these facts hold for arbitrary density matrices. 
If we now assume that the state is Gaussian, i.e., fully characterized by its CM, then the density matrix in this diagonal basis may be written as a product of thermal oscillators
\begin{equation}\label{rho_diagonal_boson}
\rho = \prod\limits_{i=1}^N Z_i\inv  e^{-\frac{1}{2} \beta_i (\tilde{p}_i^2 + \tilde{q}_i^2)}\,,
\end{equation}
where $Z_i\inv=(1- e^{-\beta_i})$ is a normalization constant while $\tilde{q}_i$ and $\tilde{p}_i$ are related to the original quadrature operators $q_i,p_i$ by means of $\tilde{\bm{X}} = M \bm{X}$. 
Moreover, the local temperatures $\beta_i$ are uniquely determined by the symplectic eigenvalues according to 
\begin{equation}\label{sigma_boson}
\sigma_i = \coth\left(\frac{\beta_i}{2}\right)\,.
\end{equation}
From Eq.~(\ref{rho_diagonal_boson}) it is now straightforward to compute the corresponding R\'enyi-$\alpha$ entropy~(\ref{renyi}), which becomes 
\begin{IEEEeqnarray}{rCl}
\label{renyi_bosonic_1}
S_\alpha(\rho) &=& \frac{1}{\alpha - 1} \sum\limits_{i=1}^N \ln \Bigg[ \left(\frac{\sigma_i+1}{2}\right)^\alpha - \left(\frac{\sigma_i-1}{2}\right)^\alpha \Bigg]\,.
\end{IEEEeqnarray}
For future convenience, let us introduce the function 
\begin{equation}\label{func_bosons}
f_\alpha^+(x) = \left(\frac{x+1}{2}\right)^\alpha - \left(\frac{x-1}{2}\right)^\alpha
\end{equation}
and rewrite \eq{renyi_bosonic_1} as
\begin{IEEEeqnarray}{rCl}
\label{renyi_bosonic_2}
S_\alpha(\rho) &=& \frac{1}{\alpha - 1} \sum\limits_{i=1}^N \ln f_\alpha^+(\sigma_i)\,.
\end{IEEEeqnarray}
From this one can already see that the case $\alpha = 2$ is rather special since $f_2^+(x)=x$, which allows expressing the R\'enyi entropy as the log determinant of the CM, namely
\begin{equation}\label{renyi_bosonic_alpha2}
S_2(\rho) = \frac{1}{2} \ln \det (\Gamma)\,.
\end{equation}
The R\'enyi entropies satisfy a monotonicity property $S_{\alpha_1}(\rho) \geq S_{\alpha_2}(\rho)$ if $\alpha_1 \leq \alpha_2$. It is therefore useful to study the limits $\alpha \to 0,\infty$. This gives rise to the max-entropy 
\begin{equation}
 S_0(\rho) = N \ln 0^+ + \sum_{i=1}^N \ln \beta_i
\end{equation}
which diverges as $N\log \alpha$ for $\alpha \to 0^+$. This is not surprising as the max-entropy is formally equivalent to the log of the rank of $\rho$ and we are working with the infinite dimensional Hilbert space for bosons. However, in suitable linear combinations of entropies such as for the (conditional) mutual information, the divergences cancel and we get a meaningful, finite, limiting result. For pure states $\beta_i \to \infty$ and $S_0(\rho) = 0$. The min-entropy can be computed as
\begin{equation}
\label{b_infty}
 S_{\infty}(\rho) = \sum_{i=1}^N \ln (1+n_i)
\end{equation}
in terms of the occupation numbers of the normal modes $n_i	=(e^{\beta_i}-1)^{-1} = (\sigma_i -1)/2$. We will see that this can be used to bound the (conditional) mutual information. 

\subsection{Fermionic systems}

We now parallel the development above for the case of fermions.
Consider a system of $N$ fermionic operators $c_i$ satisfying $\{c_i, c_j^\dagger\} = \delta_{ij}$ and $\{c_i,c_j\} = 0$. 
We define the set of majorana operators analogously to Eq.~(\ref{quadrature_bosons}), as
\begin{equation}
\gamma_{2i-1} = \frac{1}{\sqrt{2}}(c_i + c_i^\dagger)\,, 
\qquad 
\gamma_{2i} = \frac{\ii}{\sqrt{2}} (c_i^\dagger - c_i),
\end{equation}
which together form the analog of $\bm{X}$ and satisfy 
\begin{equation}\label{algebra_fermions}
\{\gamma_I,\gamma_J\} = \delta_{IJ}\,.
\end{equation}
The fermionic CM is then constructed similarly to Eq.~(\ref{boson_CM}) as 
\begin{equation}
\Gamma_{IJ} = \ii\, \langle[ \gamma_I, \gamma_J]\rangle.
\end{equation}
Any valid fermionic covariance matrix must now satisfy the bona-fide relation \cite{2009PhRvA..79a2306K}
\begin{equation}\label{bona_fide_fermion}
\ii\, \Gamma - 1 \leq 0,
\end{equation}
which again parallels Eq.~(\ref{bona_fide_boson}). This can be equivalently stated as $\Gamma\Gamma^\dagger\le1$. The Gaussian state is pure iff $\Gamma^2=-1$.

The fermionic CM can always be put in block diagonal form by an orthogonal transformation $M$ (i.e. $M M\trans = 1$) that preserves \eq{algebra_fermions}, namely
\begin{equation}\label{CMdiagonal_fermion}
M \Gamma M\trans = 
\bigoplus_{i=1}^N
\begin{pmatrix}0 &-\sigma_i\\ \sigma_i &0\end{pmatrix}\,,
\end{equation}
where $\sigma_i\in[-1,1]$. Each block is then trivially diagonalized and the eigenvalues of $\Gamma$ are simply $\pm\ii\sigma_i$. 
Unlike in the bosonic case \eq{CMdiagonal_boson}, these are the true eigenvalues of $\Gamma$. This means in particular that any matrix function $f(\Gamma)$ is block-diagonalized by the same $M$, having $f(\pm\ii\sigma_i)$ as its eigenvalues. 
A Gaussian fermionic state may then be written as 
\begin{equation}\label{rho_diagonal_fermion}
\rho = \prod\limits_{j=1}^N Z_j\inv e^{- \ii\,\beta_j \gamma_{2j-1} \gamma_{2j}},
\end{equation}
where $Z_i=2\cosh \beta_i$ and 
\begin{equation}\label{sigma_fermion}
\sigma_i = \tanh\left(\frac{\beta_i}{2}\right),
\end{equation}
which is the analog of Eq. \eq{sigma_boson}.

From this we once again can readily compute the R\'enyi entropy, which reads
\begin{IEEEeqnarray}{rCl}
\label{renyi_fermionic_1}
S_\alpha(\rho) &=& \frac{1}{1-\alpha } \sum\limits_{i=1}^N \ln \Bigg[ \left(\frac{1+\sigma_i}{2}\right)^\alpha + \left(\frac{1-\sigma_i}{2}\right)^\alpha \Bigg].
\end{IEEEeqnarray}
By defining
\begin{equation}\label{func_fermions}
f_\alpha^-(x) = \left(\frac{1+x}{2}\right)^\alpha + \left(\frac{1-x}{2}\right)^\alpha\,,
\end{equation}
we can write the R\'enyi entropy as
\begin{IEEEeqnarray}{rCl}
\label{renyi_fermionic_2}
S_\alpha(\rho) &=& \frac{1}{1-\alpha } \sum\limits_{i=1}^N \ln f_\alpha^-(\sigma_i)\,. 
\end{IEEEeqnarray}
Together with Eq. \eq{renyi_bosonic_2} this gives a fully unified description of R\'enyi entropies for both bosonic or fermionic modes. However, an important difference with respect to bosons comes from the fact that here $\pm\ii\sigma_i$ are the true eigenvalues of the CM. Namely, using $f_\alpha^-(x)=f_\alpha^-(-x)$ one can rewrite \eq{renyi_fermionic_2} as a log determinant for any $\alpha$,
\begin{IEEEeqnarray}{rCl}
\label{renyi_fermionic_3}
S_\alpha(\rho) &=& \frac{1}{2(1-\alpha)} \ln \text{det}\, f_\alpha^-(\ii\,\Gamma)\,.
\end{IEEEeqnarray}
The case $\alpha = 2$ once again allows a simple expression linear in the CM even though $f_2^-(x) = (1+x^2)/2$ is not a linear function. This is due to the peculiar structure of $\Gamma$, which implies $\det(1-\Gamma^2) = \big[\det(1+\Gamma)\big]^2$ 
and yields for the R\'enyi-2 entropy 
\begin{equation}
S_2(\rho) = N\ln(2) -  \ln \det (1+\Gamma).
\end{equation}
For fermions the max-entropy is simply
\begin{equation}
\label{f_0}
 S_0(\rho) = N\ln 2
\end{equation}
as expected from the finite dimension of the Hilbert space, except for pure states for which we should put $S_0(\rho) =0$.  The min-entropy can be expressed in exactly the same way as for bosons 
\begin{equation}
 S_{\infty}(\rho) = \sum_{i=1}^N\ln (1+n_i)	 
\end{equation}
in terms of the fermionic occupation numbers $n_i	=(e^{\beta_i}+1)^{-1} =(1-\sigma_i)/2$ of the normal modes.
%
%
\section{\label{sec:SA}Subadditivity}
%
%

In this section we study the analog of the SA inequality \eq{MI} for the R\'enyi-$\alpha$ entropies of Gaussian states, which is equivalent to non-negativity of the R\'enyi mutual information \eq{RMI}. We show that it holds for any $\alpha$ in the bosonic case, while for fermions it holds in the window $\alpha\in[0,2]$. 

\subsection{Bosons}

Let us consider a bipartition $A\cup B$ of a $N$-boson system in the Gaussian state $\rho_{AB}$. The corresponding CM can be parametrized in the block form
\begin{equation}\label{GammaAB}
\Gamma_{AB} = 
\begin{pmatrix}
\Gamma_A & \chi_{AB}  \\[0.2cm]
\chi_{AB}^T & \Gamma_B 
\end{pmatrix}\,.
\end{equation}
The reduced states $\rho_A,\rho_B$ are also Gaussian, being fully characterized by the reduced CM's $\Gamma_A,\Gamma_B$. We denote the set of symplectic eigenvalues $\sigma_i^{AB}$ of the full system by $\{d_i\}$ and collect the symplectic eigenvalues $\sigma_i^A$ and $\sigma_i^B$ of the subsystems into a single set of elements $\{c_i\}$. Both sets are assumed to be organized in non-decreasing order. Finding the necessary and sufficient conditions under which the symplectic spectra $\{c_i\}$ and $\{d_i\}$ are mutually consistent defines the Gaussian version of the so called \emph{quantum marginal problem}. These conditions have been found in \cite{2008CMaPh.280..263E} and amount to the following chain of inequalities 
\begin{subequations}\label{Eisert}
\begin{align}
\sum_{j=1}^k c_j &\ge \sum_{j=1}^k d_j\,,\qquad k=1,\ldots,N\label{Eisertsum}\\
c_n-\sum_{j=1}^{N-1}c_j &\le d_n-\sum_{j=1}^{N-1}d_j\label{Eisertdifference}\,.
\end{align}
\end{subequations}

With the conventions above, the R\'enyi mutual information \eq{RMI} associated to the R\'enyi entropies \eq{renyi_bosonic_2} can be written as
\begin{align}
\mathcal{I}_\alpha(A:B) = \sum_{j=1}^N\left[g^+_\alpha(c_j)-g^+_\alpha(d_j)\right]\,,
\end{align}
where $g^+_\alpha(x)\equiv\tfrac{1}{1-\alpha}\log f_\alpha^+(x)$. The non-negativity of $I_\alpha$ then follows straightforwardly from the fact that, for any $\alpha\ge0$ ($\alpha\ne1$), $g^+_\alpha(x)$ is a positive, monotonically increasing, concave function of $x$ in the domain $x\ge1$ (See Appendix \ref{app}). Namely,
\begin{align}\label{SA_proof_bosonic}
\mathcal{I}_\alpha &\ge\sum_{j=1}^N {g^+_\alpha}'(c_j)(c_j-d_j)\notag\\
&= \sum_{k=1}^{N-1}\!\big[{g^+_\alpha}'(c_k)-{g^+_\alpha}'(c_{k+1})\big]\!\sum_{j=1}^k\!(c_j-d_j) + {g^+_\alpha}'(c_N)\!\sum_{j=1}^N\!(c_j-d_j)\notag\\
&\ge0\,.
\end{align}
where the inequality in the first line uses concavity of $g^+_\alpha$, the second line is a convenient rewriting using Abel's partial summation formula, and the last inequality holds since each term in the previous expression is ensured to be non-negative by \eq{Eisertsum} together with monotonicity and concavity of $g^+_\alpha$. In other words, \eq{SA_proof_bosonic} shows rather remarkably the subadditivity of all the quantum R\'enyi entropies in the particular class of Gaussian states. It is interesting to note that the second constraint \eq{Eisertdifference} plays no role in the proof.

\subsection{Fermions}

Now consider a bipartite $N$-fermion Gaussian state $\rho_{AB}$ with associated fermionic CM
\begin{equation}\label{GammaABfermion}
\Gamma_{AB} = 
\begin{pmatrix}
\Gamma_A & \chi_{AB}  \\[0.2cm]
-\chi_{AB}^T & \Gamma_B 
\end{pmatrix}\,.
\end{equation}
The positive eigenvalues  $\ii\Gamma_{AB}$ by will be denoted by $\{y_i\}$, and we combine the positive eigenvalues $\ii\Gamma_A,\ii\Gamma_B$ into the set $\{x_i\}$. The ordering in the fermionic case is assumed to be non-increasing. The Sing-Thomson theorem \cite{sing1976,thompson1977,thompson1979} then implies that
\begin{align}
    \sum_{j=1}^k x_j    &\leq \sum_{j=1}^ky_j,\qquad k =1 ,\ldots, N,\\
    \sum_{j=1}^{N-1}x_j - x_N &\leq \sum_{j=1}^{N-1} y_j - y_N\,.
\end{align}
Let us define $g^-_{\alpha}(x) = (1-\alpha)^{-1}\log f^-_{\alpha}(x)$. Then we can write
\begin{equation}
 \mathcal{I}_{\alpha}(A\!:\!B) = \sum_{j=1}^N\left[g^-_{\alpha}(x_j) - g^-_{\alpha}(y_j)\right].
\end{equation}
The function $g^-_{\alpha}$ is monotonically \emph{decreasing} and concave for the interval $\alpha \in [0,2]$. Hence, for these values of $\alpha$ we can use the sequence of steps identical to Eq. \eqref{SA_proof_bosonic} to prove SA for fermionic Gaussian states and $0 \leq \alpha \leq 2$. 

For $\alpha>2$ we have checked numerically for $2$ and $3$ modes that SA is violated. Indeed, for $2$ modes it is straightforward to construct a prototypical violating CM in this range. For instance,
\begin{align}\label{fermion_2mode_violation}
\Gamma_{AB} = 
\begin{pmatrix}
0		&-\lambda	&-\xi		&0	\\
\lambda		&0		&0		&-\xi	\\
\xi		&0		&0		&-\lambda\\
0		&\xi		&\lambda	&0	
\end{pmatrix}\,
\end{align}
has $\sigma_{\pm} = |\lambda \pm \xi|$ and satisfies the bona fide condition \eq{bona_fide_fermion} as long as $|\lambda\pm \xi|\le 1$ with $|\lambda|\le1$ (for simplicity one can take both to be positive). Assuming small $\xi$ (for illustration purposes only -- this is not needed) it follows that 
\begin{align}
I_{\alpha}(A:B) &= 2g^-_{\alpha}(\lambda) - g^-_{\alpha}(\lambda+\xi) - g^-_{\alpha}(\lambda-\xi)\\
&= -{g^-_{\alpha}}''(\lambda)\,\xi^2 + \mathcal{O}(\xi^4)\,,
\end{align}
which for any $\alpha>2$ can be made to violate SA since in this case it is always possible to find a $\lambda$ for which ${g^-_\alpha}''(\lambda)$ is positive (recall that $g^-_\alpha(x)$ is no longer concave on the domain $x\in[0,1]$). A simple calculation shows that this correlation matrix $\Gamma_{AB}$ is realized by a thermal state of the Hamiltonian 
\begin{equation}
 H = (\beta_+ + \beta_-)\left[c^\dagger_1 c_1 + c^\dagger_2 c_2\right] + \ii(\beta_+ - \beta_-)\left[c^\dagger_1 c_2 - c^\dagger_2 c_1\right]\,,
\end{equation}
where $\beta_\pm=2\arctan\sigma_\pm$ and the temperature defines the unit of energy. In other words, the state $\rho\sim e^{-H}$ with the Hamiltonian above leads precisely to the SA-violating CM \eq{fermion_2mode_violation}.

%
%
\section{\label{sec:SSA}Strong subadditivity}
%
%
In this section we take a step further over Sec.~\ref{sec:SA} and present a complete survey of the regimes of validity of SSA inequality \eq{SSA} for the R\'enyi-$\alpha$ entropies of Gaussian states. 
That is, we map the full range of values of $\alpha$ for which the R\'enyi conditional mutual information \eq{RCMI} is ensured to be non-negative (SSA satisfied) and those where it is not (SSA violated). 
This is done both for bosons and fermions by numerically generating a large number of bona fide CMs and using them to find explicit violations of SSA for some $\alpha$. For bosons, we find strong evidence that SSA holds for all $\alpha\in[0,2]$ while in the fermionic case we find no violations in the interval $\alpha\in[0,\alpha_\text{max}]$ with $\alpha_\text{max}\approx 1.3$. 


\subsection{Bosons}

Consider a tripartite system in the state $\rho_{ABC}$ and let us parametrize its CM in block form as 
\begin{equation}
\Gamma_{ABC} = \begin{pmatrix}
\Gamma_A & \chi_{AB} & \chi_{AC} \\[0.2cm]
\chi_{AB}^T & \Gamma_B & \chi_{BC} \\[0.2cm]
\chi_{AC}^T & \chi_{BC}^T & \Gamma_C
\end{pmatrix}.
\end{equation}
Since the reduced density matrix is still Gaussian, the corresponding covariance matrix may be simply obtained by discarding the blocks one is tracing over. For instance, the CM $\Gamma_{AB}$ associated with $\rho_{AB}=\Tr_C\,\rho_{ABC}$ will be the one in \eq{GammaAB}.

Here we focus only on the cases where the full state $\rho_{ABC}$ is mixed, since for pure states the SSA follows from SA using the property that $S(\rho_A)=S(\rho_{\bar{A}})$ for any subsystem $A$ and its complement $\bar{A}$ \cite{Nielsen}. 

We start by reviewing the special case of $\alpha=2$ studied in \cite{Adesso2012}, the only one (apart from the trivial von Neumann case $\alpha=1)$ for which SSA is known to be satisfied. From \eq{renyi_bosonic_alpha2} it is straightforward to write the corresponding RCMI \eq{RCMI} as
\begin{equation}\label{CMI_Gaussian}
\mathcal{I}_2(A\!:\!C|B) = \frac{1}{2} \ln \frac{\det(\Gamma_{AB}) \det(\Gamma_{BC}) }{\det(\Gamma_{ABC})
\det(\Gamma_{B})}\ge0\,.
\end{equation}
The non-negativity follows immediately from the Hadamard-Fischer determinant inequality relating the minors of the symmetric positive-semidefinite matrix $\Gamma_{ABC}$.

\begin{figure}
  \includegraphics[width=0.45\textwidth]{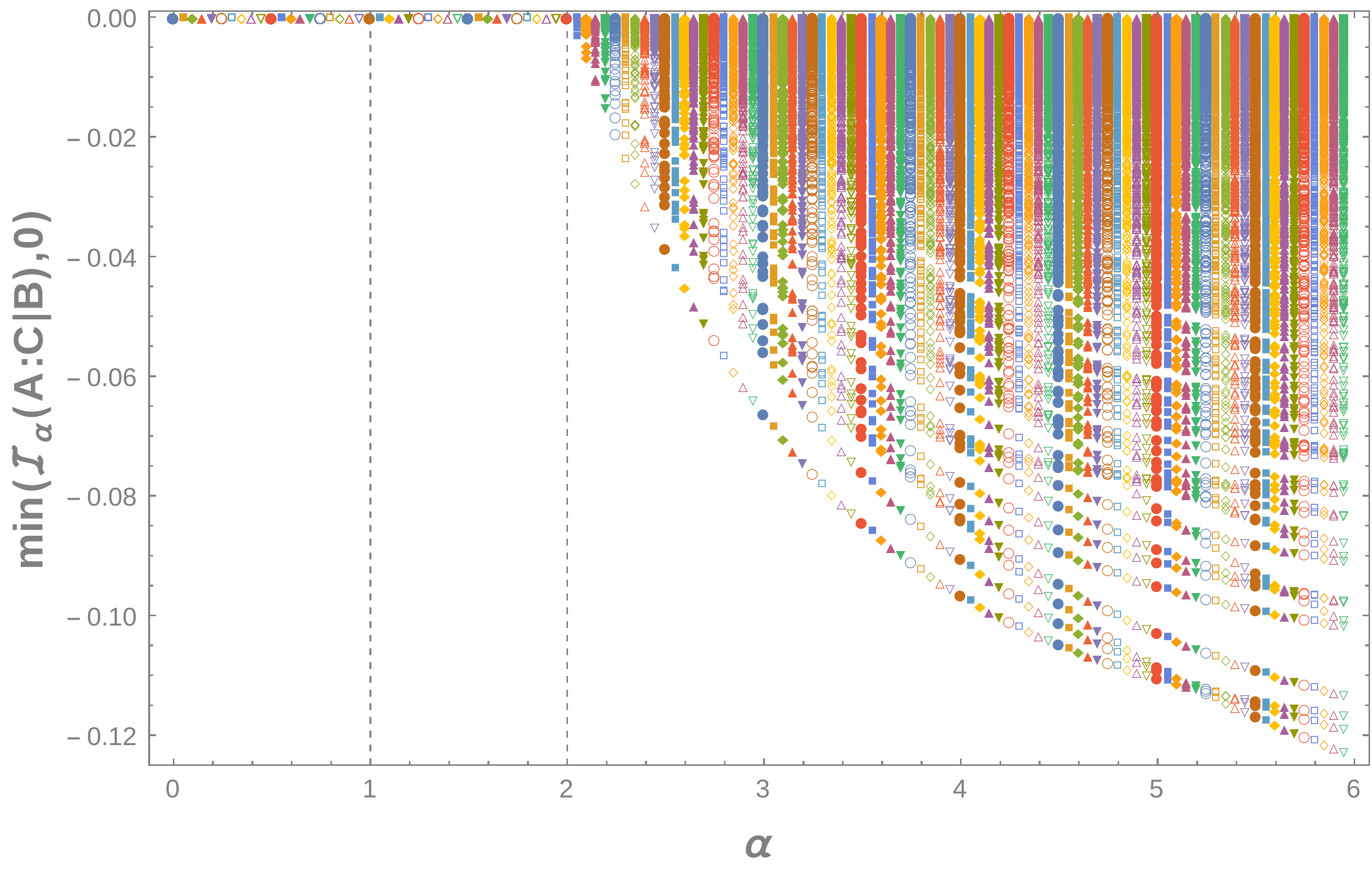}\hspace{.2cm}
  \caption{Violation of strong subadditivity (SSA) of bosonic R\'enyi entropies for three-mode Gaussian states. The plots were created by randomly generating $\sim200k$ bona fide correlation matrices and computing for different $\alpha$ the R\'enyi conditional mutual information $\mathcal{I}_{\alpha}(A\!:\!C|B)$, where $A$, $B$, and $C$ correspond to three single-mode subsystems.}
 \label{SSA_violations}
\end{figure}

For $\alpha\ne2$ we need to deal with the generic expression \eq{renyi_bosonic_1} involving particular functions of the symplectic eigenvalues. This is rather non-trivial since the only known inequalities relating the symplectic eigenvalues of the CM and those of its reduced CMs are the ones in \eq{Eisert}. We do this numerically by generating a huge number of bona fide CMs of three- and four-mode Gaussian states and use them to scan for violations of SSA by computing the RCMI \eq{RCMI} for different values of the index $\alpha$. Figure \ref{SSA_violations} shows a scatter plot of the result. It provides clear evidence that SSA is violated for all $\alpha>2$, while no violation is found for $0\le\alpha\le2$. We conjecture that this result holds true in general. Less extensive searches for SSA violations by bona fide CMs with up to six modes have not given any reason to believe that violations will be found for higher numbers of modes, but we  presently do not have a proof. 
We  hope to come back to this in future work.

Let us conclude this subsection with the statement that if $\mathcal{I}_{\infty}(A\!:\!B|C)<0$, we find that $\mathcal{I}_{\alpha}(A\!:\!B|C)<0$ for all $\alpha > \alpha_*$ with 
\begin{equation}
 \alpha_* =1 + \frac{S_{\infty}(\rho_{AB}) + S_{\infty}(\rho_{BC})}{|\mathcal{I}_{\infty}(A\!:\!B|C)|}.
\end{equation}
Hence a SSA violation for $S_{\infty}(\rho)$ implies SSA violation for any finite $\alpha$ \footnote{Using the bound $[(\sigma +1)/2 ]^{\alpha-1} \leq f^-_{\alpha}(\sigma) \leq [(\sigma +1)/2 ]^{\alpha}$ we can bound the RCMI as $S_{\infty}(\rho_{AB})\leq(\alpha-1)[\mathcal{I}_{\alpha}(A\!:\!B\!|\!C) - \mathcal{I}_{\infty}(A\!:\!B|\!C))] \leq S_{\infty}(\rho_{AB}) + S_{\infty}(\rho_{BC})$ from which the result follows.}.


\subsection{Fermions}

The story for fermions is quite similar to that of bosons. We consider a tripartite system in a state given by the block-diagonal CM 
\begin{equation}
\Gamma_{ABC} = \begin{pmatrix}
\Gamma_A & \chi_{AB} & \chi_{AC} \\[0.2cm]
-\chi_{AB}^T & \Gamma_B & \chi_{BC} \\[0.2cm]
-\chi_{AC}^T & -\chi_{BC}^T & \Gamma_C
\end{pmatrix}
\end{equation}
and its corresponding reductions to subsystems $AB,BC,A,C$. Once more we can restrict attention only to mixed states.

Even though here all the R\'enyi entropies can be written in the determinant form \eq{renyi_fermionic_3}, no claim can be made based on the Hadamard-Fisher determinant inequality since the fermionic CM is not positive semi-definite. 
In particular, unlike the case of bosons, not even the second R\'enyi entropy is guaranteed to be strongly subadditive. 

\begin{figure}
   \includegraphics[width=0.45\textwidth]{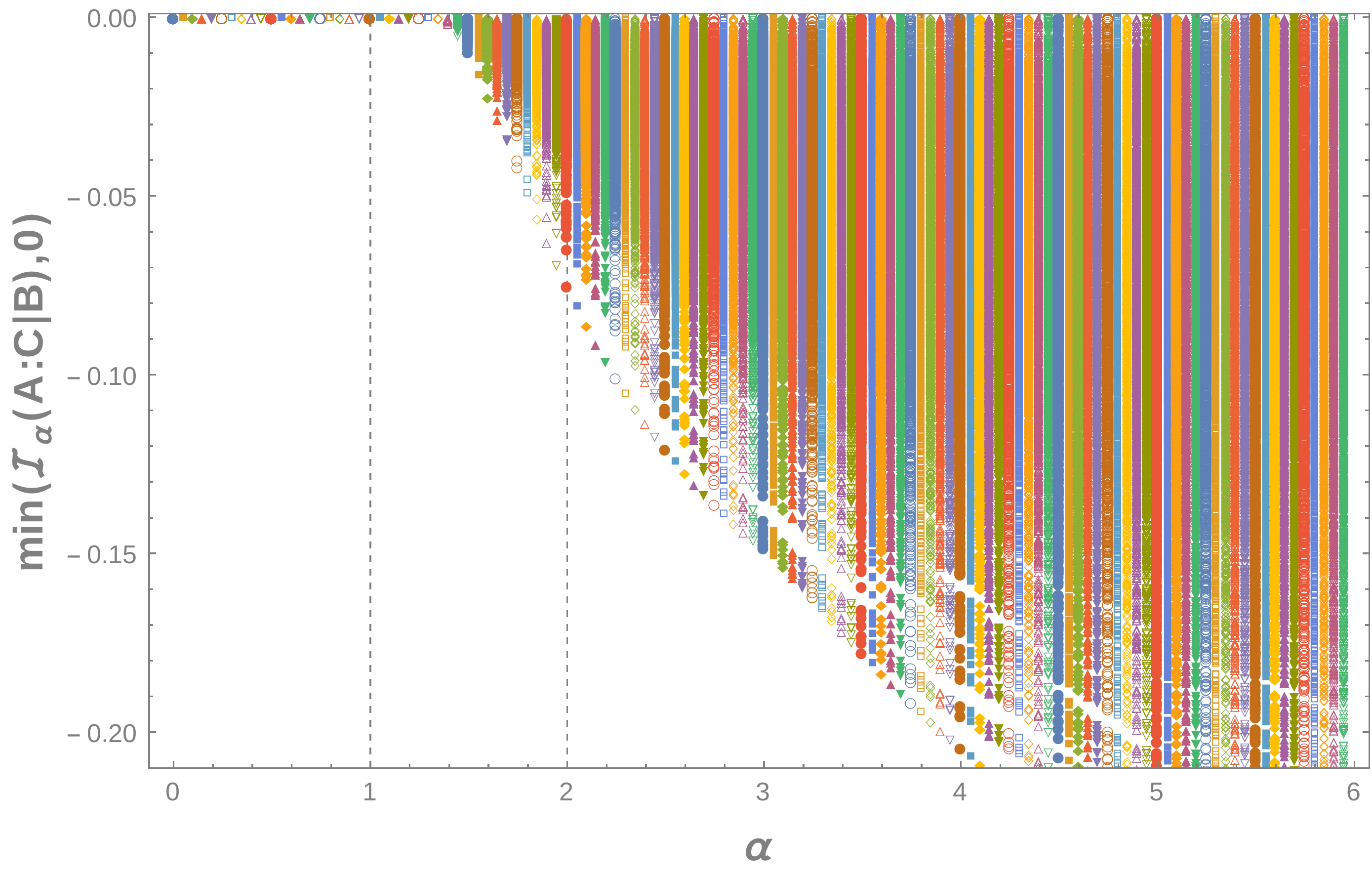}
   \caption{Violation of strong subadditivity (SSA) of fermions for three-mode Gaussian states. The plots were created by randomly generating $\sim300k$ bona fide correlation matrices and computing for different $\alpha$ the R\'enyi conditional mutual information $\mathcal{I}_{\alpha}(A\!:\!C|B),0)$ for single-mode subsystems $A$, $B$, and $C$.}
  \label{SSA_violations_fermions}
 \end{figure}

One again has to resort to numerics and deal with the generic expression \eq{renyi_bosonic_2}. We generate a large number of random bona fide CMs of three- and four-mode states and compute \eq{RCMI} looking for SSA violations as the R\'enyi index $\alpha$ is varied. The results appear in Figure \ref{SSA_violations_fermions}. We find no violations of SSA in the region $\alpha\in[0,\alpha_\text{max}]$ with $\alpha_\text{max}\approx1.3$, while many violating counterexamples are found beyond this window. The limiting value $\alpha_\text{max}$ above which violations occur is a bit surprising. The most reasonable possibility is that the limiting value is actually $\alpha = 1$, but violations for $1< \alpha  <\alpha_\text{max}$ are either very hard to find by random sampling or only possible for larger numbers of modes. We generated similarly exhaustive numbers of bona fide CMs with up to twelve modes for fermions, in an attempt to find such violations with $\alpha< \alpha_\text{max}$, but without result. Again, we lack a formal proof for SSA to hold for $\alpha \in [0,1]$, but we conjecture it to be true.

%
%
\section{\label{sec:final}Final remarks}
%
%

We have studied the R\'enyi entropies of bosonic and fermionic Gaussian states using the covariance matrix approach. 
A special effort has been made to clarify as much as possible the technical similarities between the bosonic and fermionic calculations. 
As our main result, we have obtained a complete map of the regimes of validity of the strong subaddivity (SSA) and subadditivity (SA) inequalities as a function of the R\'enyi index $\alpha$. 
We prove that SA holds for all $\alpha\ge0$ in the case of bosons and for $\alpha\in[0,2]$ in the case of fermions. The proofs rely only on concavity properties of the entropy functions together with a set of inequalities  relating the (symplectic) eigenvalues of the full correlation matrix $\Gamma_{AB}$ and those of its bipartitions $\Gamma_{A,B}$. 
The situation becomes more complicated for the SSA, for which it was necessary to resort to numerics. We provided strong numerical evidence that SSA is satisfied for $\alpha\in[0,2]$ and violated for $\alpha>2$ in the case of bosons, while for fermions we conjecture $\alpha\in[0,1]$ to be free of violations even though explicit violations are only found for $\alpha \geq \alpha_\text{max}\approx1.3$. 

Our calculations for the SA for fermions put on  firmer grounds the results reported in \cite{Kormos2017}, where it was shown that for temperature-driven quenches in the Ising model the R\'enyi mutual information in the resulting non-equilibrium steady state can become negative for $\alpha>2$ while it is definitely positive for $\alpha<2$. 
It also sheds light on the recent results of \cite{Alba:2018hie}, which showed that at late times after a quench in integrable theories the logarithmic negativity becomes proportional to the R\'enyi mutual information with $\alpha=\frac{1}{2}$. Our results guarantee that this object is always non-negative for both free bosons and free fermions, which strengthen the case for it as a good entanglement quantifier. 

As discussed in the introduction, our main goal with these results was to clarify the possible ranges of validity in which R\'enyi-based correlation quantifiers can be employed. 
This is particularly important in light of the fact that some R\'enyi entropies (particularly the R\'enyi-2) naturally appear in analytical, numerical and even experimental approaches. For instance, in Ref.~\cite{Islam2015} the authors experimentally implemented a method to measure the R\'enyi-2 entropy in a bosonic system, from which they construct the corresponding R\'enyi-2 mutual information. Their system, however, are generally in non-Gaussian states so that the positivity of the R\'enyi-2 mutual information is not guaranteed.

Notwithstanding, it is our hope that by continuing with this approach one may be able to map out these ranges of validity for different classes of states. For instance, a natural candidate would be tensor networks with well defined structures, such as matrix product states. 

An obvious continuation of this work is to prove the conjectured domains of validity of the SSA for the R\'enyi-$\alpha$ entropies. The proof is likely to involve tools other than the ones appearing in the SA proof (in particular, a $\alpha$-dependent property of the entropy functions $g^\pm_\alpha$ that restricts the proof to the range $\alpha\in[0,2]$ for bosons and $\alpha\in[0,1]$ for fermions). One can also use inspiration from standard operator-based approaches to similar proofs (as opposed to the present one based on eigenvalues), such as the one used in \cite{Adesso:2016ldo,2010arXiv1007.4626A} or the Schur complement techniques introduced in \cite{lami_log-determinant_2017,lami_schur_2016}. We hope to report on this in the near future.

\section*{Acknowledgements }

\noindent We are grateful to Diego P. Pires for helpful discussions and to Gerardo Adesso and Ludovico Lami for useful correspondence.
GTL acknowledges the International Institute of Physics, where part of this work was developed, for both the hospitality and the financial support. 
GTL also acknowledges the funding from the University of S\~ao Paulo, the S\~ao Paulo Research Foundation FAPESP (grant numbers 2016/08721-7  and 2017/20725-0), and the Brazilian funding agency CNPq (grant number INCT-IQ 246569/2014-0). GC and SE acknowledge financial support from the Brazilian ministries MEC and MCTIC.

\appendix
\section{Concavity properties of the functions $g^\pm_\alpha$}\label{app}

This appendix is devoted to study the concavity properties of the entropy functions
\begin{align}
g_\alpha^\pm(x) = \frac{\pm1}{\alpha-1}\log f^\pm_\alpha(x)\,,\quad f^\pm_\alpha(x) = \frac{[x+1]^\alpha}{2^{\alpha}}\mp\frac{[\pm (x-1)]^\alpha}{2^{\alpha}},
\end{align}
where upper signs correspond to bosons and lower signs to fermions. It will be convenient to introduce the shorthand notation $x_1=x+1$ and $x_2=\pm(x-1)$ so that both cases can be treated in a unified way as $f^\pm_\alpha(x) = 2^{-\alpha}(x_1^\alpha\mp x_2^\alpha)$. Recall that in the bosonic case the domain is $x\ge1$, meaning that $x_1\ge2$ and $x_2\ge0$; for fermions, on the other hand, the domain is $x\in[-1,1]$ but since the function is even one can focus only on the subdomain $x\in[0,1]$ so that $x_{1}\in[1,2],x_{2}\in[0,1]$. It is then straightforward to write the second derivative of $g^\pm_\alpha$ as 
\begin{align}\label{gpp}
&\partial_x^2 g^\pm_\alpha(x) = \frac{\pm1}{\alpha-1}\frac{f^\pm_\alpha \,\partial_x^2 f^\pm_{\alpha} - (\partial_x f^\pm_{\alpha})^2}{(f^\pm_\alpha)^2}\notag\\
&= \frac{\alpha(x_1x_2)^{\alpha-1}}{(\alpha-1)(f^\pm_\alpha)^2}\left[\mp\left(y^{\alpha-1}+y^{1-\alpha}\right)+(1-\alpha)(y+y^{-1})\pm2\alpha\right]\,,
\end{align}
where we introduced $y=\frac{x_1}{x_2}\ge1$. In other to prove concavity of $g_\alpha^\pm$, we have to show that $\partial_x^2 g_\alpha^\pm\le0$ for every $x$ in the domain.  

We focus first on bosons. Since the prefactor in \eq{gpp} is negative for $0\le\alpha<1$ and positive for $\alpha>1$, the task becomes to show that the term in the square brackets is non-negative in the former case and non-positive in the latter. Both results follow trivially from the Bernoulli inequalities $y^\mu \le(1-\mu)+\mu y$ (for $y>0$ and $0\le\mu\le1$) and $y^\mu \ge(1-\mu)+\mu y$ (for $y>0$ and $\mu\ge1$ or $\mu\le0$) with $\mu=\alpha$. 
This proves that $g^+_\alpha(x)$ is concave for all $\alpha\ge0$.

Now we move to fermions. For $0\le\alpha<1$, the concavity of $g^-_\alpha$ is a direct consequence of the concavity of $f^-_\alpha$ (i.e., it is preserved by the log), namely the fact that $\partial_x^2 f^-_\alpha=\alpha(\alpha-1)f^-_{\alpha-2}\le0$. For $\alpha>1$, we first notice that the prefactor in \eq{gpp} is positive and hence what remains is to show that the term inside the square brackets is non-positive. 
Clearly this is not going to happen for all $\alpha$ since the positive $y$-dependent piece can easily overcome the negative contributions for large enough $\alpha$. The limiting value for which this is avoided turns out to be $\alpha=2$. Indeed, the non-positivity of the square brackets for $\alpha\in[1,2]$ follows straightforwardly from the first Bernoulli inequality above with $\mu=\alpha-1$. In other words, $g^-_\alpha(x)$ is concave for $0\le\alpha\le2$.

\bibliography{library}

\end{document}